\documentclass[epjCONF,columns]{svjour} 
\usepackage{graphics}
\usepackage[varg]{txfonts} 
\usepackage[latin1]{inputenc}
\session-title{Model Unspecific Search in CMS}
\begin{document}

\title{Model Unspecific Search in CMS}
\author{Shivali Malhotra\inst{1}\fnmsep\thanks{\email{shivali.malhotra@cern.ch}} \and Md. Naimuddin\inst{1}\fnmsep\thanks{\email{nayeem@fnal.gov}} \and Thomas Hebbeker\inst{2} \and Arnd Meyer\inst{2} \and Holger Pieta\inst{2} \and Paul Papacz\inst{2} \and Stefan Antonius Schmitz\inst{2} \and Mark Olschewski\inst{2}}
\institute{University of Delhi, India \and III. Physikalisches Institut A, RWTH Aachen University, Germany}

\abstract{We present the results of a model independent analysis, which systematically scans the data taken by CMS for deviations from the Standard Model predictions. Due to the minimal theoretical bias this approach is sensitive to a variety of models for new physics. Events with at least one electron or muon are classified according to their content of reconstructed objects (muons, electrons, photons, jets and missing transverse energy). A broad scan of three kinematic distributions in those classes is performed by identifying deviations from Standard Model expectations, accounting for systematic uncertainties.}

\maketitle
\section{Introduction}
\label{intro}
The start-up of the LHC brings forth a new era of high energy particle physics. What we will find is yet unknown, however there is a large number of theories predicting possible outcomes.
Many of those theories are tested by dedicated analyses at CMS and the other LHC experiments. However new physics could as well manifest itself in ways no-one has yet thought of. Thus we have implemented a Model Unspecific Search in CMS: \textbf{MUSiC}.
Model unspecific searches aim to analyse a large fraction of data, systematically scanning them for deviations from the Standard Model (SM). Therefore the selection cuts are not optimized for any expected new physics signal; however, the quality of the measurement is ensured by selecting well-measured and well-understood physics objects such as isolated high-$p_{T}$ leptons. Any deviation seen might be a detector effect, a lack of understanding of the event generation and simulation or truly a new physics signal. Similar strategies have already been applied successfully at other accelerator experiments, like H1 at HERA, D$ \phi $ and CDF at Tevatron.

\section{Search Algorithm}
\label{sec:1}
New Physics usually shows up in distinctive final states and so the first task is to sort the events. We consider the following physics objects: muons ($ \mu $), electrons (e), photons ($ \gamma $), hadronic jets (jet), and missing transverse energy (MET). Events are sorted into event classes, which represent a single exclusive final state, depending on their content of reconstructed objects. To avoid using the same energy entry more than once, objects are removed if they are too close to each other: Jets are removed if there are photons or electrons nearby, and photons are removed if an electron is close. An event class can  include up to four leptons, two photons and eight jets, with at least one lepton in each event class.
After assigning events to event classes, three kinematic distributions are examined, which are promising to spot new physics:
\begin{itemize}
 \item Scalar Sum of the Transverse Momentum of all participating objects: $ \Sigma p_{T}$.
 \item Combined Invariant Mass M of those objects. For classes containing MET, the transverse invariant mass $M_{T}$ is used.
 \item  Missing Transverse Energy in classes containing MET above our predefined threshold.
\end{itemize}
Out of these distributions $ \Sigma p_{T}$ is the most general observable, sensitive to many new physics models involving heavy new particles or modified high-energy behaviour. The invariant mass allows the discovery of new resonances. New physics with heavy or highly boosted invisible particles will show up in the MET distribution.

The algorithm systematically scans for deviations, comparing the simulated SM prediction with the measured data. This is done by calculating a p-value for each connected bin region. Since all distributions are analysed as binned histograms, the bin width is adjusted to the resolution of the considered variable. It is selected to be the smallest multiple of 10GeV which exceeds the expected resolution. The p-value is the probability of a discrepancy between data and Monte Carlo, which is at least as extreme as observed, if the hypothesis is true that the Monte Carlo describes the data within its uncertainties. The region with the smallest p-value contains the most significant discrepancy and we call it the Region of Interest (\textit{RoI}).

While p denotes the probability of each RoI seen individually, it cannot be used as a statistical estimator for the global significance for such deviation in any region. A penalty factor needs to be included to account for the number of investigated regions. Using pseudo-experiments, we can compute a new probability $ \tilde{p} $ of measuring at least one deviation in any region in this distribution with a lower p-value than the one seen in the RoI of the data. The value $ \tilde{p} $ for a given distribution is then simply the number of pseudo-experiments with $p_{pseudo}$ $<$ $p_{data}$, divided by the total number of pseudo-experiments.

Both probabilities p and $ \tilde{p} $ can be translated into standard deviations of normal distributions, which is a common way to show significances. As MUSiC is sensitive to both an excess and lack of events, a two-sided normal distribution is used.

The reconstruction efficiencies and the misidentification probability are determined from simulation and uncertainties are applied to cover possible differences to data. The uncertainty on the reconstruction efficiency is estimated with 4\% for muons, 3\% for electrons and photons and 1\% for jets. For the uncertainty on the misidentification probability, we use 30\% for photons, 50\% for muons, and 100\% for electrons.

\section{Sensitivity Tests}
\label{sec:2}
To claim the absence of any obvious deviations the search needs to have a sufficient sensitivity to new physics. A number of tests scenarios has been studied to show that MUSiC would be able to successfully point out discrepancies. 
One test was to remove top-quark pair production process from the SM background and eight classes with a significance of 3$\sigma$ or higher were observed. Another test was to assume a 500 GeV Z'(neutral heavy gauge boson) as potential signal for New Physics. Here, di-muon and di-electron invariant mass distribution exceeded 4$\sigma$ level. And after assuming SUSY benchmark point LM0 as potential signal, deviations in a large number of classes were clearly visible.

\section{Results}
\label{sec:3}
The analysis has been applied to the data taken by CMS in 2010 and no significant deviations have been observed. Overall 287 distributions in 118 classes have been analysed. The distributions of the $\tilde{p}$ of all analysed event classes, separated by kinematic distribution, i.e. Scalar Sum of the Transverse Momentum (Fig. 1), Invariant Mass (Fig. 2) and Missing Transverse Energy (Fig. 3) are shown here. The shaded area is the SM simulation and the crosses represent the data, and no unexpected deviations are found. The complete results for all the classes can be found at: \textit{https://cms-project-music.web.cern.ch/cms-project-music}.

\begin{figure}

\resizebox{0.75\columnwidth}{!}{%
  \includegraphics{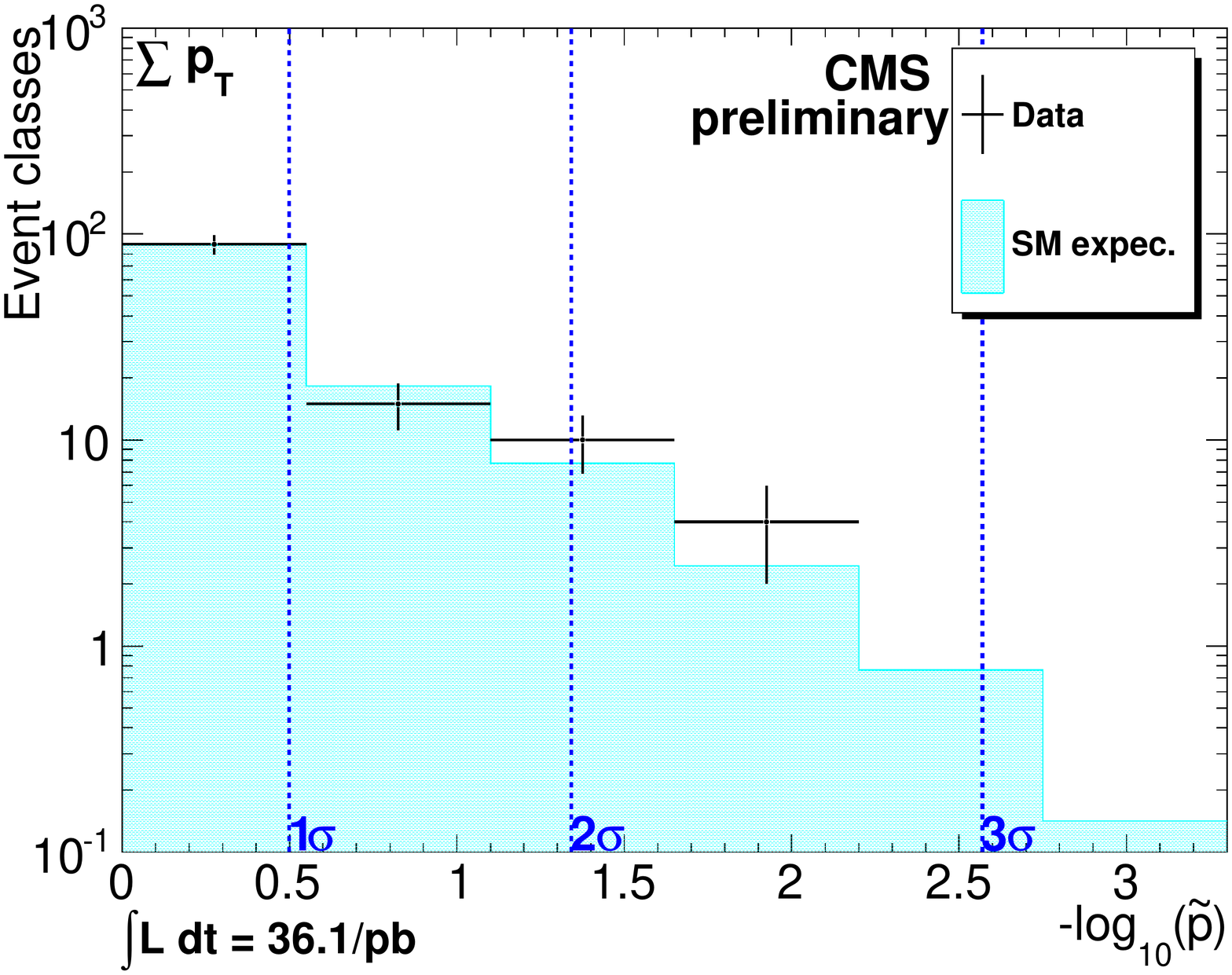} }
\caption{Sum pT distribution of $\tilde{p}$ of data-vs-SM (black crosses) compared to SM-vs-SM (shaded area)}
\label{fig:1}       
\end{figure}

\begin{figure}

\resizebox{0.75\columnwidth}{!}{%
  \includegraphics{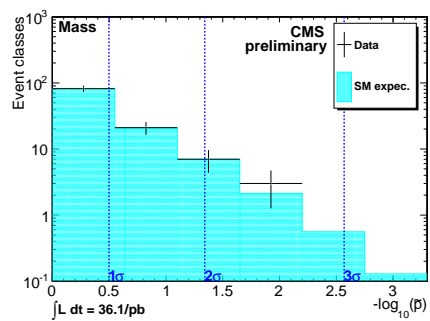} }
\caption{Invariant Mass distribution of $\tilde{p}$ of data-vs-SM (black crosses) compared to SM-vs-SM (shaded area)}
\label{fig:2}       
\end{figure}

\begin{figure}

\resizebox{0.75\columnwidth}{!}{%
  \includegraphics{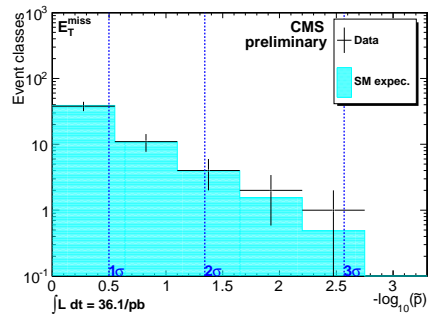} }
\caption{Missing Transverse Energy distribution of $\tilde{p}$ of data-vs-SM (black crosses) compared to SM-vs-SM (shaded area)}
\label{fig:3}       
\end{figure}

\section{Summary}
\label{sec:4}

The implementation and results of a model independent analysis of the 2010 CMS data (at an integrated luminosity of 36.1/pb) have been presented. Studies have been performed to evaluate the sensitivity of the approach to certain kinds of scenarios. MUSiC illustrates good agreement between data and SM expectation within uncertainties.

\end{document}